\documentclass[iop]{emulateapj-rtx4}

\usepackage{mathrsfs }
\usepackage{natbib}
\usepackage{amssymb}
\usepackage{apjfonts}
\usepackage{ulem} 
\usepackage{graphicx}
\begin{document}

\title{The MOSDEF Survey: Neon as a Probe of ISM Physical Conditions at High Redshift\altaffilmark{1}}

\author{
Moon-Seong Jeong,\altaffilmark{2}
Alice E. Shapley,\altaffilmark{2}
 Ryan L. Sanders,\altaffilmark{3,4}
 Jordan N. Runco,\altaffilmark{2}
 Michael W. Topping,\altaffilmark{2}
 Naveen A. Reddy,\altaffilmark{5}
 Mariska Kriek,\altaffilmark{6}
 Alison L. Coil,\altaffilmark{7}
 Bahram Mobasher,\altaffilmark{5}
 Brian Siana,\altaffilmark{5}
 Irene Shivaei,\altaffilmark{8,4}
 William R. Freeman,\altaffilmark{5}
 Mojegan Azadi,\altaffilmark{9}
 Sedona H. Price,\altaffilmark{10}
 Gene C. K. Leung,\altaffilmark{7}
 Tara Fetherolf,\altaffilmark{5}
 Laura de Groot,\altaffilmark{11}
 Tom Zick,\altaffilmark{6}
 Francesca M. Fornasini,\altaffilmark{9}
 Guillermo Barro\altaffilmark{12}
}

\altaffiltext{1}{Based on data obtained at the W.M. Keck Observatory, which is operated as a scientific partnership among the California Institute of Technology, the University of California,  and the National Aeronautics and Space Administration, and was made possible by the generous financial support of the W.M. Keck Foundation.}
\altaffiltext{2}{Department of Physics and Astronomy, University of California, Los Angeles, 430 Portola Plaza, Los Angeles, CA 90095, USA}
\altaffiltext{3}{Department of Physics, University of California, Davis, 1 Shields Avenue, Davis, CA 95616, USA}
\altaffiltext{4}{Hubble Fellow}
\altaffiltext{5}{Department of Physics and Astronomy, University of California, Riverside, 900 University Avenue, Riverside, CA 92521, USA}
\altaffiltext{6}{Astronomy Department, University of California at Berkeley, Berkeley, CA 94720, USA}
\altaffiltext{7}{Center for Astrophysics and Space Sciences, Department of Physics, University of California, San Diego, 9500 Gilman Drive., La Jolla, CA 92093, USA}
\altaffiltext{8}{Steward Observatory, University of Arizona, 933 N Cherry Ave, Tucson, AZ 85721, USA}
\altaffiltext{9}{Harvard-Smithsonian Center for Astrophysics, 60 Garden Street, Cambridge, MA, 02138, USA}
\altaffiltext{10}{Max-Planck-Institut f\"ur Extraterrestrische Physik, Postfach 1312, Garching, 85741, Germany}
\altaffiltext{11}{Department of Physics, The College of Wooster, 1189 Beall Avenue, Wooster, OH 44691, USA}
\altaffiltext{12}{Department of Phyics, University of the Pacific, 3601 Pacific Ave, Stockton, CA 95211, USA}
\email{aes@astro.ucla.edu}

\shortauthors{Jeong et al.}



\shorttitle{Neon and the ISM at High Redshift}

\begin{abstract}
We present results on the properties of neon emission
in $z\sim2$ star-forming galaxies drawn from the 
MOSFIRE Deep Evolution Field (MOSDEF) survey. Doubly-ionized
neon ([NeIII]$\lambda3869$) is detected at $\geq3\sigma$
in 61 galaxies, representing $\sim$$25$\% of the MOSDEF sample with
H$\alpha$, H$\beta$, and [OIII]$\lambda5007$ detections at
similar redshifts. We consider the neon emission-line properties of both individual galaxies
with [NeIII]$\lambda3869$ detections and composite $z\sim2$ spectra binned by stellar mass.
With no requirement of [NeIII]$\lambda3869$ detection,
the latter provide a more representative picture of neon emission-line
properties in the MOSDEF sample. The [NeIII]$\lambda3869$/[OII]$\lambda3727$
ratio (Ne3O2) is anti-correlated with stellar mass in $z\sim2$ galaxies, as expected based on the
mass-metallicity relation. It is also positively correlated
with the [OIII]$\lambda5007$/[OII]$\lambda3727$ ratio (O32),
but $z\sim2$ line ratios are offset towards higher Ne3O2 at fixed O32, compared
with both local star-forming galaxies and individual H~II regions. 
Despite the offset towards higher Ne3O2 at fixed O32 at $z\sim2$, biases in inferred Ne3O2-based metallicity
are small. Accordingly, Ne3O2 may serve as an important metallicity indicator deep into the
reionization epoch.  Analyzing
additional rest-optical line ratios including [NeIII]$\lambda3869$/[OIII]$\lambda5007$ (Ne3O3)
and [OIII]$\lambda5007$/H$\beta$ (O3H$\beta$), we conclude that the nebular emission-line
ratios of $z\sim2$ star-forming galaxies suggest a harder ionizing spectrum
(lower stellar metallicity, i.e., Fe/H) at fixed gas-phase oxygen abundance, compared to systems at $z\sim0$. 
These new results based
on neon lend support to the physical picture
painted by oxygen, nitrogen, hydrogen, and sulfur emission, of an ionized ISM in high-redshift
star-forming galaxies irradiated by chemically young, $\alpha$-enhanced massive stars.
\end{abstract}


\section{Introduction}
\label{sec:intro}
Rest-optical emission-line spectroscopy provides a comprehensive
view of the ionized phase of the interstellar medium (ISM)
in star-forming galaxies. Key quantities inferred
from nebular emission lines include
the instantaneous star-formation rate (SFR),
dust content, gas-phase oxygen abundance, virial
and non-virial kinematics, and physical ISM gas density.
These measurements also describe the
radiation field exciting interstellar gas, its intensity
and spectral shape, and whether its source is primarily massive
stars or an active galactic nucleus (AGN).
With the advent of sensitive multi-object near-IR
spectrographs on 8-10-meter-class ground-based telescopes,
there has been much recent progress 
assembling rest-optical emission-line measurements for
star-forming galaxies during the epoch of peak
star formation in the universe ($z\sim1.5-3.5$).
The next frontier consists of extending such measurements
into the epoch of reionization with the {\it James Webb
Space Telescope} ({\it JWST}).

Typically, the strongest rest-optical emission features are
the lower Balmer lines (H$\alpha$ and H$\beta$) and transitions
from singly- and doubly-ionized oxygen ([OII]$\lambda\lambda3726,3729$, hereafter [OII]$\lambda3727$,
and [OIII]$\lambda\lambda4959,5007$). These features, along
with those from singly-ionized nitrogen ([NII]$\lambda6584$)
and sulfur ([SII]$\lambda\lambda6717,6731$), have been used to infer
the nature of the ionized ISM at $z\sim2$ and the important
ways in which its properties are distinct from those
at $z\sim0$ \citep[e.g.,][]{steidel2014,shapley2019}. There is emerging consensus based on recent
work from the Keck Baryonic Structure Survey \citep[KBSS;][]{steidel2014} and
MOSFIRE Deep Evolution Field (MOSDEF) Survey \citep{kriek2015} that
the pattern of rest-optical emission lines observed in
$z\sim2$ star-forming galaxies reflects their chemically young
stellar populations. Massive stars in such galaxies
are enhanced in $\alpha$ elements (e.g., oxygen) relative to Fe,
and, accordingly, the ionizing spectrum is systematically harder
in these distant galaxies compared to local galaxies with similar gas-phase chemical
abundances \citep{steidel2016,strom2017,shapley2019}.

The [NeIII]$\lambda3869$ feature provides
another important probe of the ionized ISM.  \citet{nagao2006} proposed
the ratio [NeIII]$\lambda3869$/[OII]$\lambda3727$ (Ne3O2)
as an empirical metallicity indicator \citep[see also][]{shi2007,
perezmontero2007}, due to its monotonic anti-correlation with
oxygen abundance.  As discussed in \citet{levesque2014},
this anti-correlation arises because [NeIII]$\lambda3869$/[OII]$\lambda3727$,
like [OIII]$\lambda5007$/[OII]$\lambda3727$ (O32), reflects
the ionization parameter, which is anti-correlated with metallicity. Due to the difference
in ionization potential for [NeIII]$\lambda3869$ and
[OIII]$\lambda5007$, however, a comparison of Ne3O2
and O32 ratios in galaxies can also provide a window into the
spectral shape of the ionizing radiation field \citep{strom2017}.
Ne3O2 has the advantage over O32 that it is insensitive
to dust extinction, due to the proximity in wavelength
of [NeIII]$\lambda3869$ and [OII]$\lambda3727$. Also,
based on the short wavelengths of its component features,
Ne3O2 can be measured from the ground to $z\sim5$ \citep{shapley2017}
and, with {\it JWST}, all the way to $z\sim12$.

[NeIII]$\lambda3869$ is typically fainter than the strongest
lines from hydrogen, oxygen, nitrogen and sulfur, and is only
typically detected in lower-metallicity, higher-excitation
sources. Accordingly, samples of individual high-redshift
galaxies with [NeIII] measurements are not representative of
the star-forming galaxy population as a whole. \citet{zeimann2015}
used stacked low-resolution near-IR grism
spectra from the {\it Hubble Space Telescope} ({\it HST}) to investigate
the typical properties of neon in low-mass (median $10^9 \mbox{M}_{\odot}$)
$z \sim2$ star-forming galaxies. Here, we 
use higher-resolution and deeper near-IR spectroscopy
from the MOSDEF survey to investigate the properties
of both individual $z\sim2$ galaxies with [NeIII] detections, as well
as the average neon emission-line properties in our sample as a
function of stellar mass ($\mbox{M}_*$). These observations provide important complementary constraints
on the nature of the ionizing spectrum in high-redshift star-forming
galaxies, and have implications for inferring gas-phase metallicities
at the earliest times. In Section~\ref{sec:obs}, we present our observations
and samples. Section~\ref{sec:results} contains our main results, while Section~\ref{sec:discussion}
contains a discussion of the implications of these results.
Throughout, we assume a $\Lambda$CDM cosmology with
$H_0=70$ km s$^{-1}$ Mpc$^{-1}$, $\Omega_m=0.3$, and $\Omega_{\lambda}=0.7$.
In addition to Ne3O2 and O32, defined previously, we adopt
the abbreviations Ne3O3 for the line ratio [NeIII]$\lambda3869$/[OIII]$\lambda5007$, and O3H$\beta$
for [OIII]$\lambda5007$/H$\beta$.

\begin{figure}[ht!]
\centering
\includegraphics[width=\linewidth]{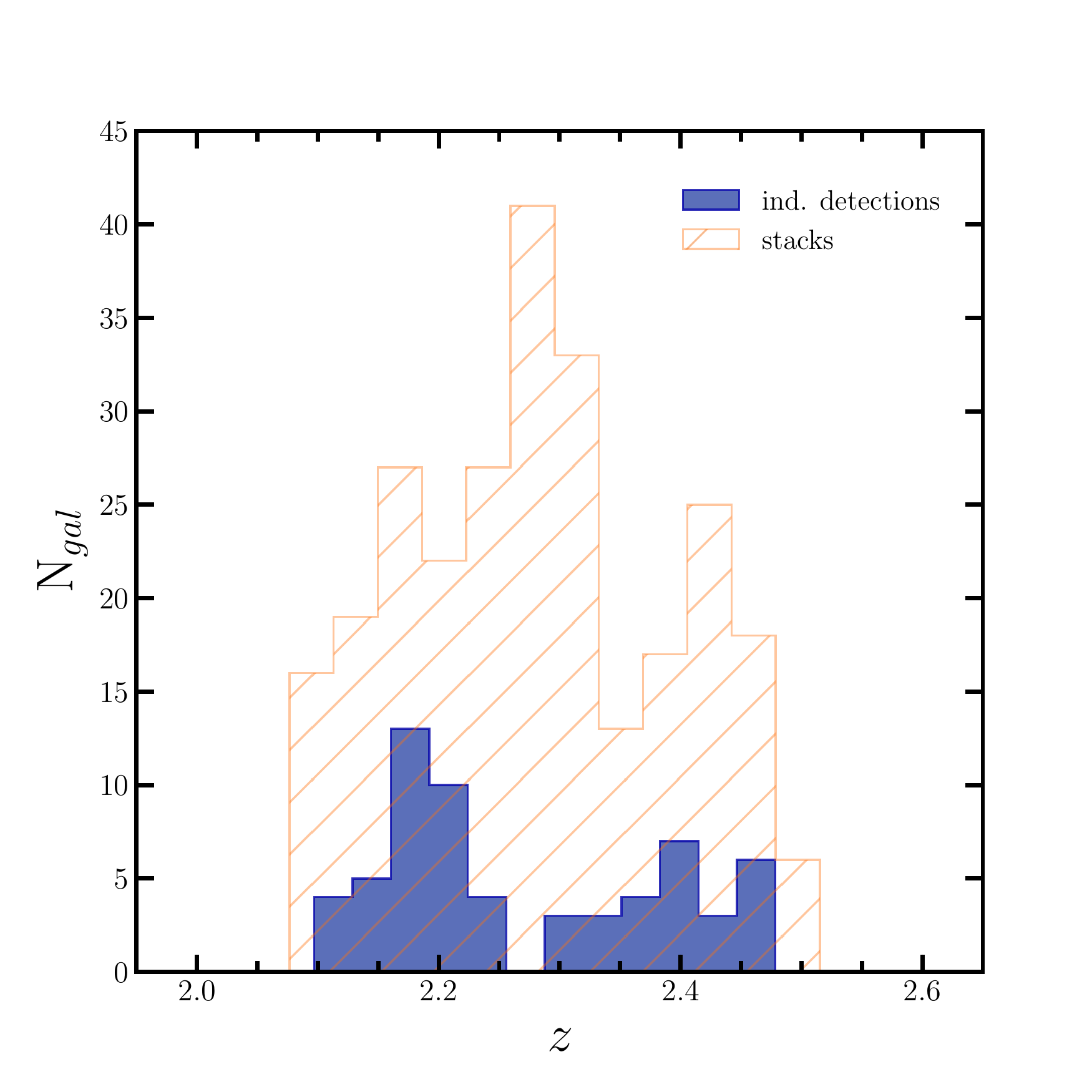}
\caption{Redshift histograms for $z\sim2$ MOSDEF galaxies with individual
[NeIII] detections (blue, solid) and those included in the larger sample of
MOSDEF composite spectra (orange, hatched). The median redshift for the
sample of [NeIII] detections is $z_{{\rm med,[NeIII]}}=2.223$, while
that for the composite sample is $z_{{\rm med,comp}}=2.29$.}
\label{fig:zhist}
\end{figure}

\section{Observations and Samples}
\label{sec:obs}

\subsection{MOSDEF survey} 
\label{sec:obs-mosdef} 
Our analysis of emission-line ratios at high redshift focuses on a sample of 
galaxies drawn from the MOSDEF survey \citep{kriek2015}. MOSDEF is a near-IR 
spectroscopic survey of $\sim1500$ galaxies at $1.37 \leq z \leq 3.80$. These 
galaxies lie in the well-studied CANDELS fields (AEGIS, COSMOS, GOODS-N, 
GOODS-S, and UDS) for which there exist ample supplementary multi-wavelength 
data.
With MOSDEF, we used the Multi-object Spectrometer for Infrared Exploration 
\citep[MOSFIRE;][]{mclean2012} on the Keck~I telescope to obtain moderate-resolution ($R\approx 
3000-3650$) $J$-, $H$-, and $K$-band spectra probing the rest-optical 
($\sim3700-7000$\AA) for target galaxies. Further details of the survey 
observations and data reduction are found in \citet{kriek2015}. 
Here we discuss some of the aspects of MOSDEF that are most relevant to our 
analysis.

\begin{figure*}[ht!]
\centering
\includegraphics[width=0.9\linewidth]{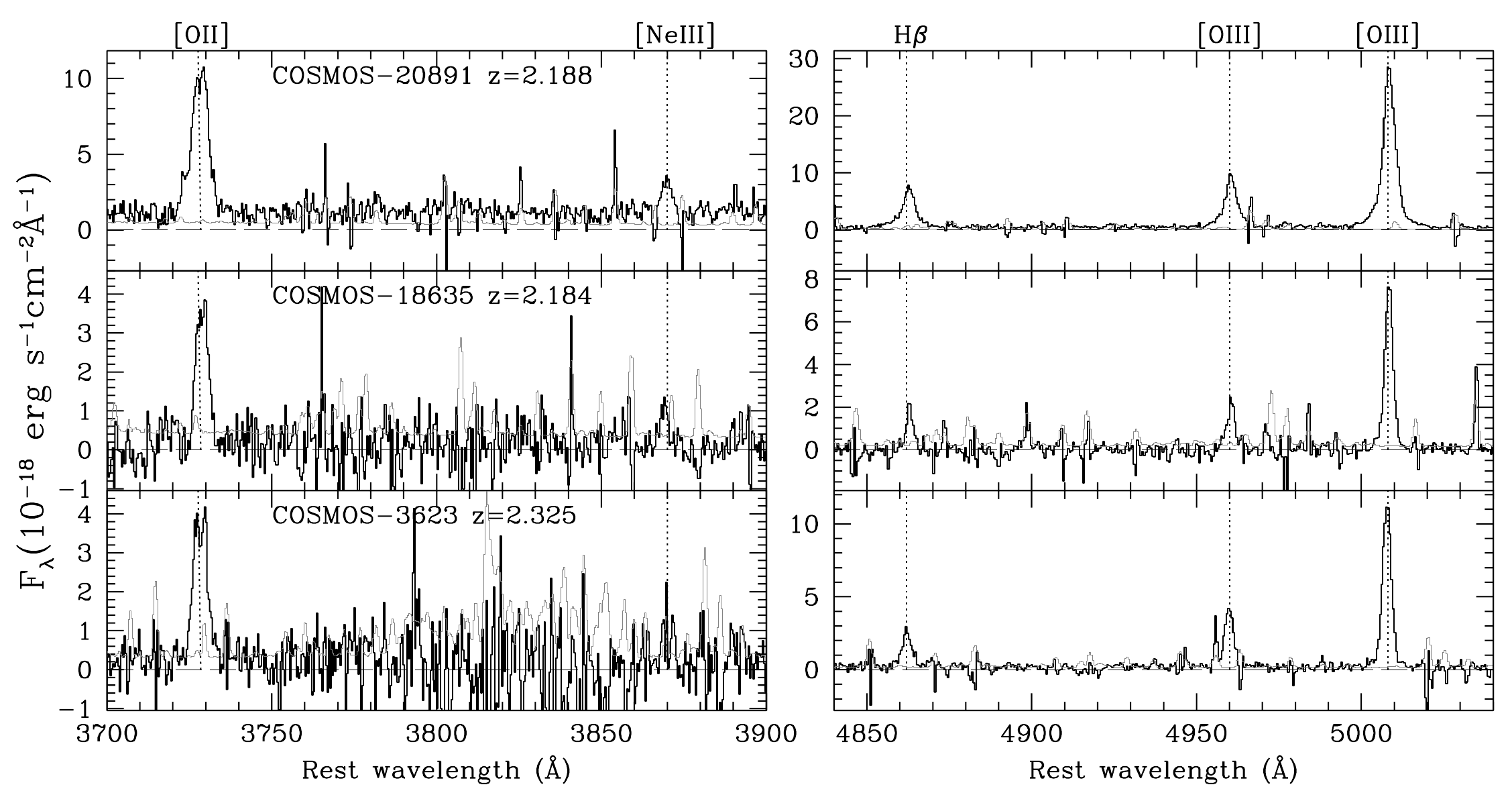}
\caption{Flux-calibrated $J$-band (left) and $H$-band (right) spectra for
three examples from our sample of 61 galaxies with individual [NeIII]
detections, demonstrating the range of [NeIII] S/N in the sample.
Key nebular emission lines are labeled.
$J$-band covers [OII] and [NeIII], while $H$-band covers
H$\beta$ and [OIII]. From top to bottom, we show COSMOS-20891, with
[NeIII] S/N$\sim12.5$ (the highest [NeIII] S/N in the sample),
COSMOS-18635, with the median [NeIII] S/N $\sim4$, and COSMOS-3623,
at the threshold of [NeIII] detection (S/N$\sim3$).}
\label{fig:jhspec}
\end{figure*}

We studied the rest-optical emission-line fluxes, stellar masses and dust 
reddening estimated for $z\sim2$ MOSDEF galaxies. Stellar masses were 
estimated by fitting emission-line corrected photometry obtained from the 
3D-HST photometric catalogs \citep{skelton2014, momcheva2016} with FAST, an 
SED-fitting code \citep{kriek2009}. The \citet{conroy2009} flexible stellar 
population synthesis (SPS) model, a \citet{chabrier2003} initial mass 
function (IMF) and the \citet{calzetti2000} dust attenuation curve were 
assumed. We estimated nebular reddening, $E(B-V)_{{\rm neb}}$, using the 
stellar-absorption-corrected H$\alpha$/H$\beta$ Balmer decrement
for galaxies with H$\alpha$ and H$\beta$ measurements 
with signal-to-noise ratio (S/N) greater than or equal to three (S/N $\geq3$). 
We assumed an intrinsic H$\alpha$/H$\beta$ emission-line ratio of 2.86 
and the \citet{cardelli1989} extinction law for our calculations.

MOSDEF galaxies were targeted in three redshift bins -- $1.37\leq z\leq1.70$, 
$2.09\leq z \leq2.61$, and $2.95\leq z \leq3.80$ -- to ensure 
coverage of bright rest-frame optical emission lines within windows of 
atmospheric transmission \citep{kriek2015}. Here we select star-forming 
galaxies in the $2.09\leq z \leq2.61$ range with robust detection (S/N$\geq3$) in the nebular 
emission lines of interest: [OII]$\lambda$3727, [NeIII]$\lambda$3869, H$\beta$, 
[OIII]$\lambda5007$, and H$\alpha$. We selected this redshift bin 
because all of the emission lines of interest fall into the $J$, $H$, and $K$ 
bands, which are covered by the MOSDEF survey.  We also excluded AGNs flagged 
on the basis of X-ray emission, {\it Spitzer}/IRAC colors, or [NII]/H$\alpha$ 
ratios $\geq 0.5$ \citep{coil2015,azadi2017}. 

Among the 547 star-forming MOSDEF galaxies with a measured redshift at 
$2.09\leq z\leq2.61$, only 61 satisfy the requirements for inclusion 
in the neon detection sample. The redshift distribution of individual 
[NeIII] detections is shown in Figure~\ref{fig:zhist}.
The median stellar mass and dust-corrected H$\alpha$ SFR of the 
galaxies with [NeIII] detections are, respectively, $\log(\mbox{M}_*/\mbox{M}_{\odot})=9.83$ 
and 26$\mbox{M}_{\odot}\mbox{ yr}^{-1}$. In Figure~\ref{fig:jhspec}, we show 
examples of $J$- and $H$-band spectra for galaxies with [NeIII] detections. 
These spectra cover [OII] and [NeIII] in $J$, and H$\beta$ and [OIII] in $H$, and demonstrate the range of S/N spanned by the sample of individual [NeIII] detections. 

Given that such a small fraction of the $z\sim2$
MOSDEF sample shows individual [NeIII] detections, we also analyzed the 
median emission-line ratios of a more representative sample based on 
composite spectra in four bins of stellar mass. This sample of 264 galaxies 
within the same redshift range only has the requirement of H$\alpha$ and 
H$\beta$ detections, along with excluding AGN. Composite spectra were 
constructed as in \citet{sanders2018}, and the mass ranges, median mass, number of galaxies,
and measured line ratios of each composite spectrum are listed in Table~\ref{tab:composite}. 
The redshift distribution of the 
stacked sample is also shown in Figure~\ref{fig:zhist}.
Requiring coverage of [NeIII]$\lambda3869$ for both individual detections
and composite spectra reduces the redshift range to $2.09\leq z\leq2.50$.

\begin{deluxetable*}{ccccccc}[b!]
\tabletypesize{\footnotesize}
\tablecolumns{7}
\tablewidth{7in}
\tablecaption{Masses and Emission-line Ratios of $z\sim2$ MOSDEF stacks\label{tab:composite}}
\tablehead{
\colhead{$\log\left(\frac{M_*}{M_{\odot}}\right)$\tablenotemark{a}} & \colhead{$\log\left(\frac{M_*}{M_{\odot}}\right)_{\rm med}$\tablenotemark{b}} &
\colhead{$N_{\rm gal}$\tablenotemark{c}} & \colhead{Ne3O2\tablenotemark{d}} & \colhead{Ne3O3\tablenotemark{d}} & \colhead{O3H$\beta$\tablenotemark{d}} & \colhead{O32\tablenotemark{d}}
\vspace{.10cm}
}
\startdata
    8.23--9.51 & 9.29 & 66 & $-0.73^{+0.04}_{-0.04}$ & $-0.99^{+0.04}_{-0.04}$ & $0.63^{+0.02}_{-0.02}$ & $0.26^{+0.02}_{-0.02}$ \\[0.10cm]
    9.52--9.83 & 9.72 & 66 & $-0.90^{+0.05}_{-0.05}$ & $-0.93^{+0.05}_{-0.05}$ & $0.53^{+0.02}_{-0.02}$ & $0.03^{+0.02}_{-0.02}$ \\[0.10cm]
    9.84--10.19 & 9.99 & 66 & $-1.19^{+0.08}_{-0.10}$ & $-1.02^{+0.08}_{-0.10}$ & $0.39^{+0.02}_{-0.02}$ & $-0.17^{+0.02}_{-0.01}$ \\[0.10cm]
    10.20--11.27 & 10.43 & 66 & $-1.04^{+0.06}_{-0.07}$ & $-0.70^{+0.06}_{-0.07}$ & $0.28^{+0.02}_{-0.02}$ & $-0.34^{+0.02}_{-0.02}$
\enddata
\tablenotetext{a}{Range of $\log(M_*/M_{\odot})$ of galaxies in a bin.}
\tablenotetext{b}{Median $\log(M_*/M_{\odot})$ of galaxies in a bin.}
\tablenotetext{c}{Number of galaxies in a bin.}
\tablenotetext{d}{Dust-corrected emission-line ratios measured from stacked spectra in each bin.}
\end{deluxetable*}

\subsection{Local comparison samples}
\label{sec:obs-local}

We compare our $z\sim2.3$ MOSDEF sample to local datasets 
including archival data from  the 
Sloan Digital Sky Survey (SDSS) DR7 \citep{abazajian2009}. 
We obtain stellar masses and emission-line properties for individual 
SDSS galaxies at $0.04\leq z \leq0.10$ from the MPA-JHU DR7 
release of spectrum measurements.\footnote{https://wwwmpa.mpa-garching.mpg.de/SDSS/DR7/}
We also made use of the stacked SDSS measurements in bins of 
stellar mass from \citet{andrews2013}. Finally, as in 
\citet{sanders2017}, we use a local H~II region catalog 
from \citet{pilyugin2016}, supplemented by observations 
from \citet{croxall2016} and \citet{toribio2016}, for a 
total of 1050 H~II regions. Out of the $\sim97,000$ 
SDSS galaxies with $3\sigma$ detections of [OII], H$\beta$, [OIII], 
H$\alpha$, and [NII], which fall on the star-forming side of the 
AGN-SF curve of \citet{kauffmann2003}, only 6615 have 
corresponding detections of the typically weaker [NeIII] feature. 
On the other hand, [NeIII] is detected in all of the \citet{andrews2013} 
stacks at $\log(\mbox{M}_*/\mbox{M}_{\odot})=8.0-11.0$ and in 930 out of 1050 of the 
H~II regions.

\section{Results}
\label{sec:results}
\subsection{Ne3O2 and Stellar Mass}
\label{sec:res-ne3o2m}
It has been shown that $z\sim2$ galaxies have systematically higher 
ionization parameters (in an empirical sense, O32 values)
at fixed stellar mass relative to local 
galaxies \citep{strom2017,sanders2016}. 
Furthermore, \citet{sanders2016} and \citet{sanders2018} find
an anti-correlation between O32 and stellar
mass in $z\sim2$ MOSDEF galaxies, as observed in the local universe. 
This anti-correlation
is consistent with the existence of the mass-metallicity relationship at 
both low  and high redshift, given the well-known
anti-correlation between ionization parameter and nebular 
metallicity \citep{perezmontero2014}.

\begin{figure}[ht!]
\centering
\includegraphics[width=\linewidth]{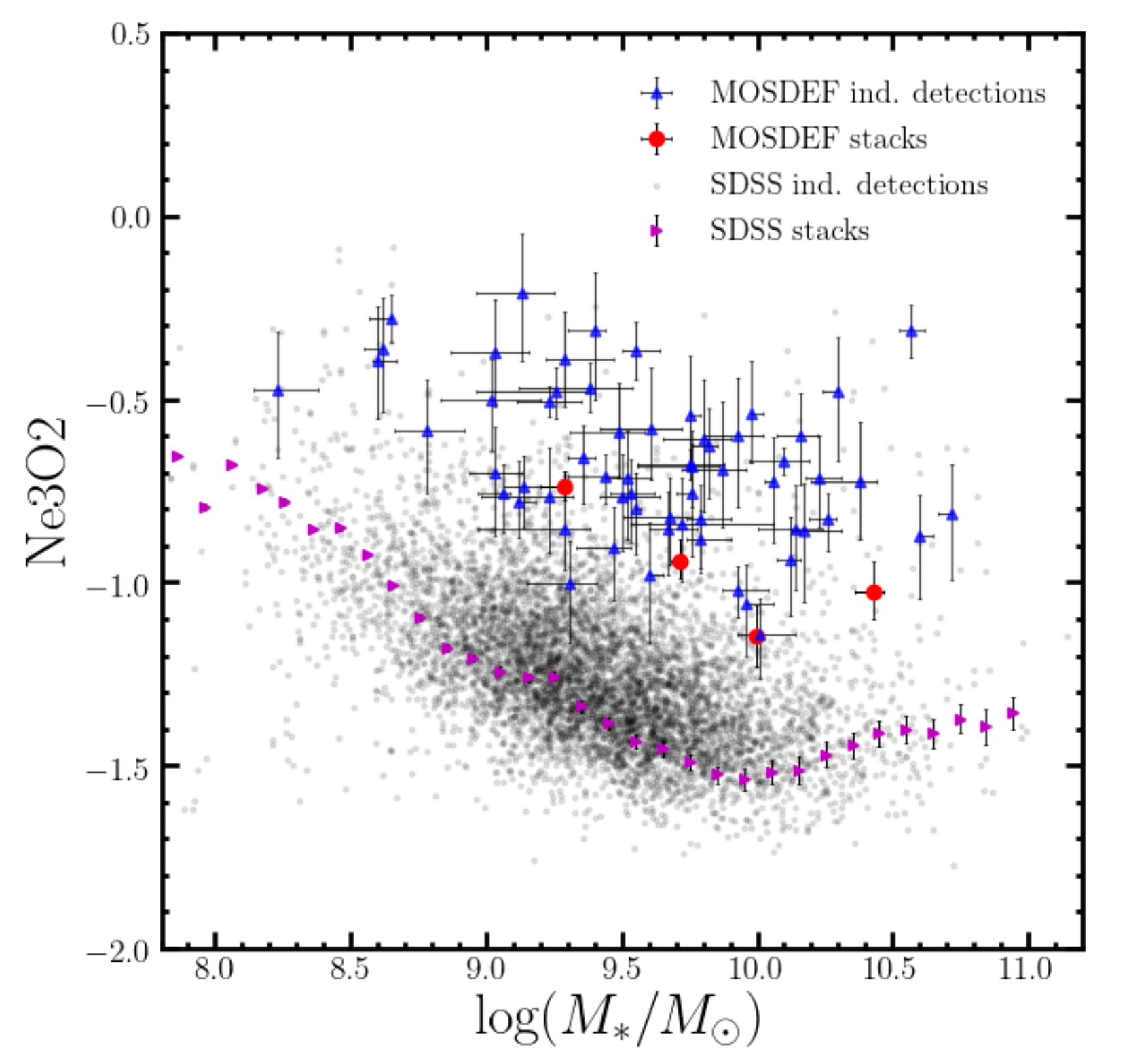}
\caption{Ne3O2 vs. stellar mass for individual $z\sim2$ MOSDEF
galaxies (blue triangles) and stacked data (red circles) compared with
local SDSS galaxies (small grey circles) and stacked data from
\citet{andrews2013} (purple triangles). In both local and $z\sim2$
datasets, the stacked Ne3O2 values fall below the median of the
individual points at fixed stellar mass, indicating the bias in the
[NeIII]-detected samples.}
\label{fig:ne3o2mass}
\end{figure}

We explored the relationship between Ne3O2 and $\mbox{M}_*$ in the MOSDEF 
sample, for both individual [NeIII] detections and stacked spectra in 
bins of $\mbox{M}_*$. In Figure~\ref{fig:ne3o2mass},
a clear anti-correlation is observed between Ne3O2 and $\mbox{M}_*$, 
although the stacked MOSDEF points fall below the median of the 
individual [NeIII] detections. A similar difference
in Ne3O2 is observed between the individual [NeIII]-detected 
SDSS points and the stacked spectra
from \citet{andrews2013}. Such offsets are expected
given the incompleteness of the [NeIII]-detected samples, 
and underscores the importance
of utilizing the full dataset through stacking analysis when the 
goal is to characterize global emission-line trends and their 
corresponding physical scaling relationships
such as the mass-metallicity relation \citep{sanders2018}.  At fixed stellar mass, 
$z\sim2$ MOSDEF stacked spectra are $\sim0.5$~dex higher
in Ne3O2 than their local stacked counterparts from \citet{andrews2013}.
We note that the highest-mass (lowest-O32) MOSDEF bin deviates towards 
elevated Ne3O2 from the monotonic
trend present in the three lower-mass bins, which may indicate low-level 
AGN activity that is not detected on an individual basis.

\subsection{Neon Line Ratios as a Probe of ISM Conditions}
\label{sec:res-ne3emlines}
\begin{figure*}[ht!]
\centering
\includegraphics[width=\linewidth]{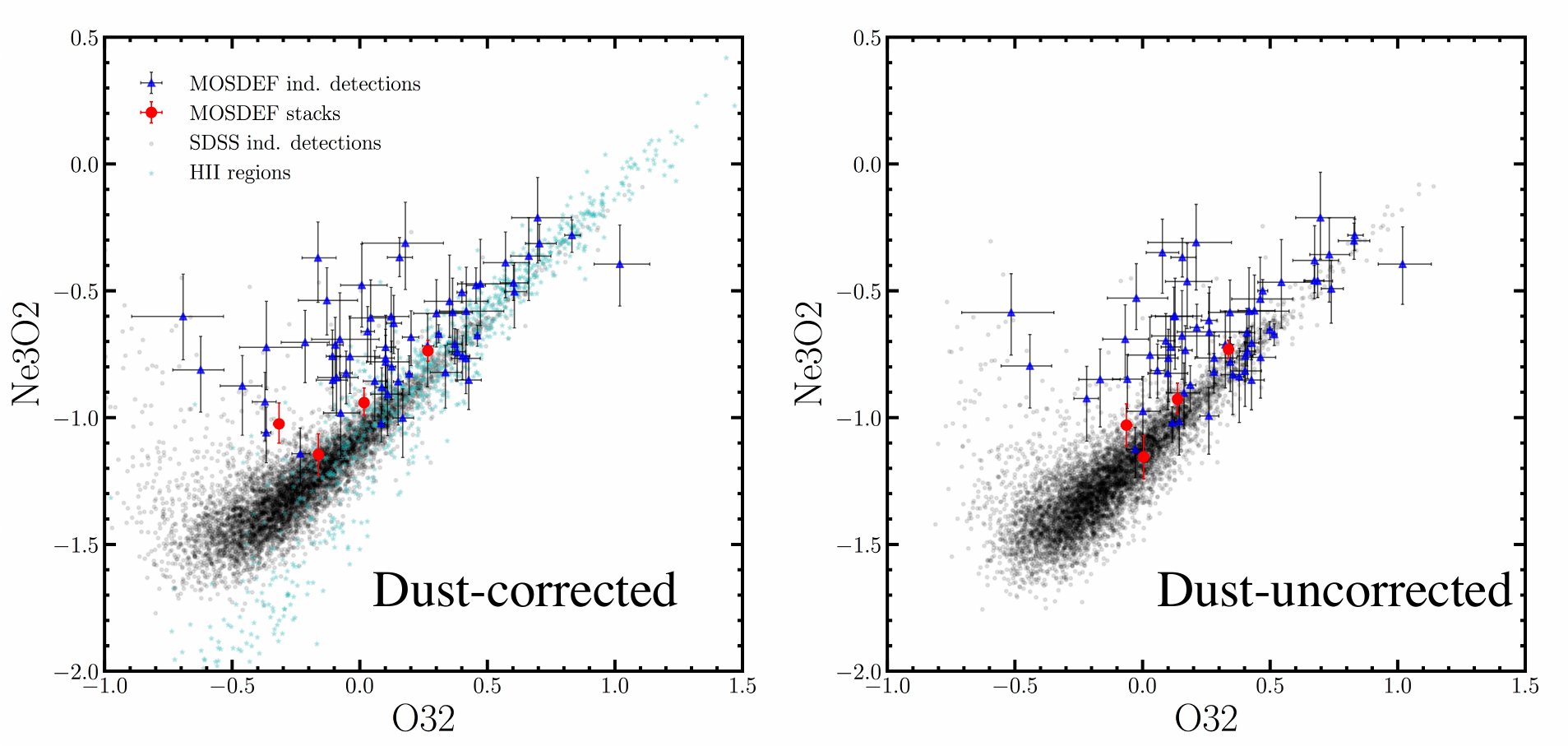}
\caption{\textbf{Left}: The distribution of dust-corrected Ne3O2 vs.
O32 for $z\sim2$ MOSDEF individual galaxies (blue triangles) and
spectral stacks in bins of stellar mass (red circles), along with
local SDSS galaxies (small grey circles), and H~II regions
(small cyan stars). MOSDEF galaxies are offset towards larger Ne3O2
at fixed O32, relative to both SDSS galaxies and H~II regions.
\textbf{Right}: Ne3O2 vs. O32 for MOSDEF galaxies and stacks, and local SDSS galaxies, with no dust-correction applied. H~II regions
are not included here as only dust-corrected values were tabulated. The same positive offset in Ne3O2 at fixed O32 is observed with no dust correction.}
\label{fig:ne3o2o32}
\end{figure*}

The line ratio Ne3O2 provides a useful probe of the 
ionization parameter, and, accordingly, the nebular oxygen
abundance in star-forming regions \citep[e.g.,][]{nagao2006,levesque2014}. 
As demonstrated by \citet{strom2017},
the combination of Ne3O2 and other line ratios such as O32 can also be 
used to trace quantities such as the hardness
of the ionizing spectrum (i.e., the metallicity of the massive stars 
providing the ionizing radiation field).
Due to the higher ionization energy of Ne${^+}$ (40.96~eV) 
compared with that of O$^{+}$ (35.12~eV), harder ionizing
spectra at fixed nebular metallicity yield larger Ne3O2 ratios at 
fixed O32 ratio. In Figure~\ref{fig:ne3o2o32},
we show the relationship between Ne3O2 and O32 for $z\sim2$ MOSDEF 
individual galaxies and stellar mass stacks,
individual SDSS galaxies with [NeIII] detections, and local H~II regions. 
In this space, the SDSS galaxies
and H~II regions diverge at low O32, likely as the contribution from 
diffuse ionized gas (DIG) becomes
more significant in the integrated galaxy spectra 
\citep{sanders2017,shapley2019}. Regardless of which
local comparison sample is used, the $z\sim2$ MOSDEF galaxies, 
including the representative
$z\sim2$ MOSDEF stacks, are offset from local systems towards 
higher Ne3O2 at fixed O32. Given the wavelength
spacing of the individual features in O32, it is necessary to 
dust correct this feature to infer the intrinsic
line ratios in the absence of dust attenuation (left-hand panel of 
Figure~\ref{fig:ne3o2o32}). 
Recent MOSDEF observations suggest that the \citet{cardelli1989}
curve is appropriate for high-redshift nebular dust corrections (Reddy et al. 2020), 
yet, to rule out the possibility that
a biased dust correction at high redshift introduces the Ne3O2 offset, 
we also plot the Ne3O2 vs. O32 relation {\it uncorrected} for dust (right-hand panel).
Even without dust correction, the $z\sim2$ points are offset towards higher 
Ne3O2 at fixed O32.

\begin{figure*}[h]
\centering
\includegraphics[width=\linewidth]{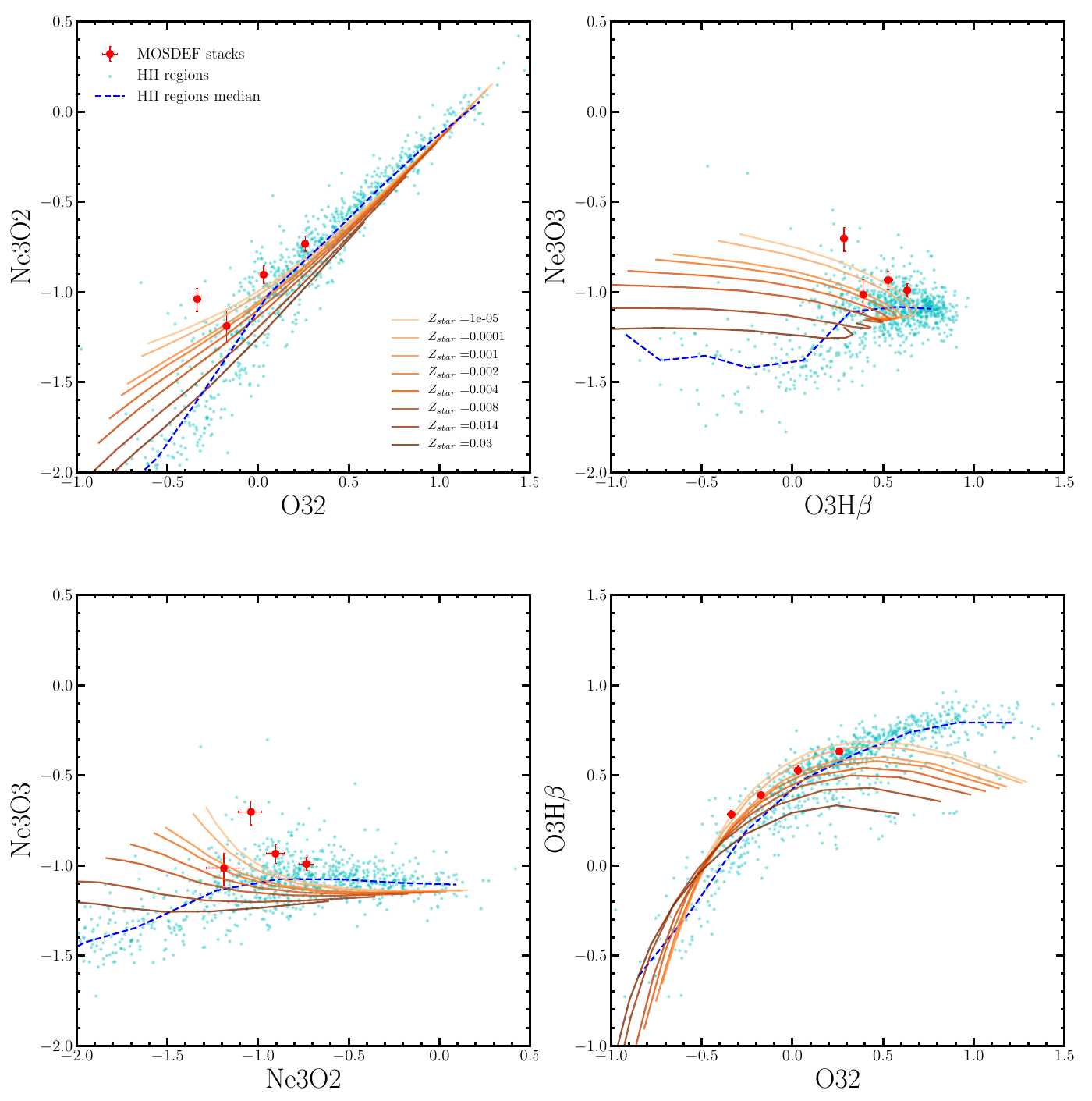}
\caption{Projections of the dust-corrected
emission-line ratio space of [OII]$\lambda3727$, [NeIII]$\lambda3869$, H$\beta$, and [OIII]$\lambda5007$. In each panel, $z\sim2$ MOSDEF
stacks are indicated with red circles and local H~II regions with small cyan stars. Blue dashed curves indicate the running median line ratios for the H~II regions. Cloudy+BPASS photoionization model curves are 
overplotted. Each model curve represents a different stellar metallicity (ionizing spectrum), along which  ionization parameter ($U$)
and nebular metallicity ($12+\log(\mbox{O/H})$) are tied together as described in the text. For each curve of fixed
stellar metallicity, $U$ increases and $12+\log(\mbox{O/H})$ decreases with increasing O32 or Ne3O2. 
In the plot of Ne3O3 vs. O3H$\beta$,
ionization parameter increases (nebular metallicity decreases) starting at the upper left of each curve. 
}
\label{fig:ne3-emline-panels}
\end{figure*}

As discussed in \citet{shapley2019}, given the $\sim$ two orders of magnitude higher typical SFR surface
densities in $z\sim2$ galaxies compared with their $z\sim0$ star-forming galaxy counterparts, the integrated
spectra from these $z\sim2$ galaxies likely lack a significant contribution from DIG. Accordingly, the most appropriate
local comparison sample for $z\sim2$ star-forming galaxies is actually individual H~II regions, as opposed to SDSS 
galaxies, the latter of which have significant DIG contribution that leads to systematically different emission-line sequences
\citep{sanders2017}.
In order to gain physical insight into the observed Ne3O2 offset at fixed O32, we compare $z\sim2$ MOSDEF stacked
line ratios with local H~II regions and photoionization model curves in Figure~\ref{fig:ne3-emline-panels}. We consider not only the space
of Ne3O2 vs. O32, but also Ne3O3 vs. O3H$\beta$, Ne3O3 vs. Ne3O2, and O3H$\beta$ vs. O32, 
to untangle the various physical properties in play. 

We use the code Cloudy \citep[v17.01;][]{ferland2017} with input ionizing spectra
from the Binary Population And Spectral Synthesis (BPASS) v2.2.1 models \citep{eldridge2017,stanway2018}. For the BPASS models, 
we followed the methodology of \citet{topping2020}, varying stellar metallicity over a wide range from extremely subsolar
to supersolar, assuming a \citet{chabrier2003} IMF with a high-mass cutoff of $100\mbox{M}_{\odot}$ and a constant star-formation history.
We also fixed the model age to be $t=10^{8.6}$~yr, although the results are not sensitive to age for $t>10^7$~yr.
We furthermore collapsed the model grid to reflect the observed anti-correlation between ionization parameter ($U$)
and oxygen abundance ($12+\log(\mbox{O/H})$) at $z\sim0$ \citep{perezmontero2014}, 
normalized to pass through the $U$ and $12+\log(\mbox{O/H})$
values found for $z\sim2$ galaxies in \citet{topping2020}: $\log(U) = -1.06 \times (12+\log(\mbox{O/H}))+5.78$.

In all panels of Figure~\ref{fig:ne3-emline-panels} that include neon emission-line ratios (Ne3O2 vs. O32, Ne3O3 vs. O3H$\beta$,
and Ne3O3 vs. Ne3O2), we observe that $z\sim2$ galaxies are offset from the median $z\sim0$ H~II region distribution
towards higher Ne3O2 at fixed O32, higher Ne3O3 at fixed O3H$\beta$, and higher Ne3O3 at fixed Ne3O2. The distribution of 
model curves in these diagnostic diagrams suggests that such offsets are indicative of a harder ionizing spectrum
(lower stellar metallicity) at fixed nebular metallicity for $z\sim2$ galaxies. In contrast, changes in ionization
parameter at fixed nebular metallicity \citet{bian2020} correspond to shifts {\it along} the distribution of datapoints.
Quantitatively, it is not straightforward to assign a stellar metallicity to the $z\sim2$ galaxies based
on these diagrams, as the model curves 
do not overlap the full distribution of H~II regions and $z\sim2$ galaxies. However,
the negative gradient in stellar metallicity as model curves shift in the direction of the $z\sim2$ emission-line offset 
provides a consistent explanation of the differences between high-redshift and local galaxies. We additionally note
that $z\sim2$ galaxies are offset towards higher O3H$\beta$ at fixed O32, which also
suggests a lower stellar metallicity at fixed $12+\log(\mbox{O/H})$ for high-redshift galaxies.

\section{Discussion}
\label{sec:discussion}

The most comprehensive previous analysis of nebular neon emission
at $z\sim2$ was performed by \citet{zeimann2015}. Based on composite
{\it HST} grism spectra for a sample of 236 galaxies, these authors
found a significant enhancement of Ne3O3 relative to local galaxies. Ne3O2
was also $\sim+0.2$~dex more offset at fixed O32 in this sample, relative
to our findings. A likely explanation for the smaller enhancements we
observe in Ne3O2 and Ne3O3 is contamination of [NeIII] by a blend of He~I and H$\zeta$ at $\lambda=3889$\AA\ in
low-resolution {\it HST} grism spectra (Zeimann 2020, private communication),
but not in the significantly higher-spectral-resolution MOSDEF stacks, in which
He~I+H$\zeta$ is clearly detected and distinct from [NeIII]. Our results also build on those
from \citet{strom2017}, who showed that the average spectrum of 30 $z\sim2$ KBSS galaxies
in the Ne3O2 vs. O32 diagram was consistent with a harder ionizing spectrum
at fixed nebular metallicity, relative to local systems. 
We now show a monotonic trend in Ne3O2 vs. O32 at $z\sim2$ in bins of
stellar mass over the range $10^9-10^{10}\mbox{M}_{\odot}$ -- but one that is still
offset towards enhanced Ne3O2 relative to the local sequence.

The observed enhancements in Ne3O2 and Ne3O3 at $z\sim2$ relative to local star-forming
regions have implications for interpreting other emission-line properties at high redshift.
There has been much debate in the literature regarding the observed properties of $z\sim2$
galaxies in the space of [OIII]$\lambda5007$/H$\beta$ vs. [NII]$\lambda6584$/H$\alpha$.
Possible explanations for the offset between high-redshift and local galaxies include a higher ionization parameter
or harder ionizing spectrum, a different N/O abundance pattern, higher electron density,
or previously unrecognized AGN activity \citep[e.g.,][]{kashino2017,steidel2016,sanders2016,kewley2013,
wright2010}. \citet{shapley2019} used observations of [SII] line ratios in $z\sim2$ MOSDEF
galaxies to show that a harder ionizing spectrum at fixed nebular metallicity provides the best explanation
of the sulfur emission-line properties in $z\sim2$ MOSDEF galaxies. Our results based on neon support the interpretation
of a harder ionizing spectrum at fixed nebular metallicity. The shape of the ionizing spectrum 
is primarily modulated by the Fe abundance of the massive stars providing the ionizing radiation --
with harder spectra corresponding to lower Fe abundance  -- and the nebular metallicity
traces the abundance of $\alpha$ elements such as oxygen. Thus, a harder ionizing spectrum at fixed
nebular metallicity is indicative of $\alpha$-enhancement (i.e., super-solar O/Fe values)
in the massive stars exciting the ionized gas in star-forming regions at $z\sim2$ \citep[e.g.,][]{steidel2016,sanders2020}.
Such abundance patterns are a natural consequence of the young median stellar population
ages in $z\sim2$ star-forming galaxies \citep{topping2020}. Given the strong evidence for
$\alpha$-enhancement in distant star-forming galaxies, $\alpha$-enhanced chemical abundance
patterns should be considered when modeling their stellar populations and ionized gas.

An important related question is whether Ne3O2 can be used as a robust metallicity indicator
at high redshift, given the inferred differences in the ionizing spectrum presented here.
At $z\leq3$ there are several other line ratios (e.g., O32, 
O3H$\beta$, $\mbox{R23}\equiv\log(([\mbox{OII}]\lambda3727+[\mbox{OIII}]\lambda\lambda4959,5007)/\mbox{H}\beta$, $\mbox{N2}\equiv [\mbox{NII}]\lambda6584/\mbox{H}\alpha$,
O3N2$\equiv \mbox{O3H}\beta$/N2) that can be used for inferring gas-phase oxygen abundance.
However, at $z=3.8-5.0$ from the ground (i.e., $K$-band), and $z\sim9-12$ (i.e., F290LP filter on {\it JWST}/NIRSPEC)
from space, Ne3O2 is the only rest-optical line ratio available. We turn to the ``local analog'' sample of \citet{bian2018},
which serves as an appropriate calibration dataset for $z\sim2$ star-forming galaxies \citep{sanders2020}, 
in the absence of a statistical sample of high-redshift objects with direct $T_e$-based metallicities
from which an independent empirical calibration of Ne3O2 to $12+\log(\mbox{O/H})$ can be performed.
The $z\sim2$ MOSDEF stacked data is on average 0.08~dex higher in Ne3O2 at fixed O32, relative
to the \citet{bian2018} stacks. The local analog calibration between Ne3O2 and  $12+\log(\mbox{O/H})$
is $12+\log(\mbox{O/H})=7.8-0.63\times \mbox{Ne3O2}$. Therefore, the average enhancement in Ne3O2 at $z\sim2$ corresponds
to an average underestimate in metallicity of only $\sim0.05$~dex, relative to the local analog sample.
Such differences are small compared to the uncertainties in inferred metallicity. Accordingly, we conclude 
that any biases resulting from estimating metallicity from Ne3O2 alone are not significant at $z\sim2$. Based
on calibration with appropriate lower-redshift analogs, 
this line ratio may provide extremely important insights into chemical enrichment deep into the reionization
era ($z>9$).

\section*{Acknowledgements}
We acknowledge support from NSF AAG grants AST-1312780, 1312547, 1312764, and 1313171, grant AR-13907
from the Space Telescope Science Institute, and grant NNX16AF54G from the NASA ADAP program.
We also acknowledge a NASA contract supporting the ``WFIRST Extragalactic Potential Observations (EXPO)
Science Investigation Team" (15-WFIRST15-0004), administered by GSFC.
We thank the 3D-HST collaboration, who provided us with spectroscopic
and photometric catalogs used to select MOSDEF targets and derive
stellar population parameters.
We benefitted from useful conversations with Greg Zeimann.
We finally wish to extend special thanks to those of Hawaiian ancestry on
whose sacred mountain we are privileged to be guests.


\end{document}